\documentstyle[12pt,preprint,tighten,aps]{revtex}
\newtheorem{theo}{Theorem}

\newtheorem{prop}{Proposition}

\def\B{{\tilde{B}}}
\def\ut#1{\rlap{\lower1ex\hbox{$\sim$}}{#1}}
\def\contr#1{\raise1ex\hbox{\small $[#1]$}}
\def\H{{\cal H}}
\def\SO{{\rm SO}}

\preprint{\vbox{\baselineskip=12pt
\rightline{CGPG-99/1-1}}}

\begin{document}

\title{BF Description of Higher-Dimensional Gravity Theories}
\author {L.\ Freidel${}^{1,2}$, K.\ Krasnov${}^1$ and
R.\ Puzio${}^1$\thanks{E-mail addresses: freidel,krasnov,puzio@phys.psu.edu}}
\address{1. Center for Gravitational Physics and Geometry \\
Department of Physics, The Pennsylvania State University \\
University Park, PA 16802, USA}

\address{2. Laboratoire de Physique  \\
Ecole Normale Sup\'erieure de Lyon \\
46, all\'ee d'Italie, 69364 Lyon Cedex 07, France}
\date{\today}
\maketitle

\begin{abstract}

It is well known that, in the first-order formalism, pure three-dimensional
gravity is just the BF theory. Similarly, four-dimensional general relativity
can be formulated as BF theory with an additional constraint term added to
the Lagrangian. In this paper we show that the same is true also for 
higher-dimensional Einstein gravity: in any dimension gravity can be 
described as a constrained BF theory. Moreover, in any dimension these constraints
are quadratic in the B field. After describing in details the structure 
of these constraints, we scketch the ``spin foam'' quantization of these
theories, which proves to be quite similar to the spin foam quantization 
of general relativity in three and four dimensions. In particular, in 
any dimension, we solve the quantum constraints and find the so-called 
simple representations  and  intertwiners. These exhibit a simple 
and beautiful structure that is common to all dimensions.

\end{abstract}

\newpage

\section{Introduction}
\label{sec:Intr}

In three spacetime dimensions Einstein's general relativity becomes a
beautiful and simple theory. There are no local degrees of
freedom, and gravity is an example of topological field
theory. Owing to this fact, a variety of techniques from TQFT
can be used, and a great deal is known about
quantization of the theory.
More precisely, when written in the first order formalism,
three-dimensional gravity is just the BF theory, whose action is
given by:
\begin{equation}
S_{\rm BF} = \int_{\cal M} {\rm Tr} (B\wedge F).
\end{equation}
Here $\cal M$ is the spacetime manifold, $F$ is the curvature of
the spin connection, and $B$ is the frame field one-form.
The trace is taken in the Lie algebra of the relevant gauge group,
which in the case of 3D is
given by $\rm SO(2,1)$  for Lorentzian spacetimes and by $\SO (3)$
in the Euclidean case. The quantization of BF theory is well-understood,
both canonically and by the path integral method, at least in the
Euclidean case. This is one of
the possible ways to construct
quantum gravity in three spacetime dimensions:
it exists as a topological field theory.\footnote{%
Let us note, however, that there are subtle problems with large
gauge transformations and degenerate metrics in this theory,
which, to our knowledge, are not yet resolved in a completely
satisfactory manner. See, for instance, \cite{3DVolume} for a
discussion of these issues.}

It is tempting to apply the beautiful quantization
methods from TQFT to other, more
complicated theories, including those with local
degrees of freedom. An interesting proposal along these
lines was made in a series of papers by Martellini and
collaborators \cite{Martellini}, who proposed to treat
Yang-Mills theory as a certain deformation of the BF theory.
This gives an interesting
picture of the confining phase of Yang-Mills theory.

Recently, a  proposal was made suggesting a way to
apply the ideas and methods from TQFT to four-dimensional gravity.
The new approach to quantum gravity, for which the
name ``spin foam approach'' was proposed in \cite{Baez},
lies on the intersection between  TQFT and loop quantum gravity
(see \cite{LQG} for a recent review on ``loop'' gravity).
As it was advocated by Rovelli and Reisenberger \cite{Rov},
the results of the ``loop'' approach suggest
a possibility of constructing the partition function
of $4$D gravity as a ``spin foam'' model. The first
``spin foam'' model of $4$D gravity was constructed by
Reisenberger \cite{R}, and was intimately related to the self-dual canonical
(loop) quantum gravity.
Later, Barrett, Crane \cite{BC} and Baez \cite{Baez} proposed another model
based on the study of the geometry of a 4-simplex. Both
spin foam models deeply use the fact that
Einstein's theory in four dimensions
can be rewritten as a BF theory  with additional quadratic
constraints.
It is a constrained $\rm SU(2)$ BF theory
\cite{Plebanski,CDJ} in the self-dual case,
and a constrained  $\rm SO(4)$  BF theory \cite{Laurent} for the
Barrett, Baez and Crane model.
In both cases, the resulting
quantum model is given by a certain deformation of the
topological BF theory.

While the approach of Martellini et al. \cite{Martellini},
which treats Yang-Mills theory as a deformation of the BF
theory, is clearly not limited to any spacetime dimension, one might
suspect that the similar strategy in the case of gravity
works only in three and four dimensions. Indeed, it is
believed that in order to quantize a theory
in a way similar to the one used in TQFT,
the theory must at least have the property that its phase
space consists of pairs connection -- conjugate electric field.
However, already in the case of four dimensions, the fact that the
gravitational phase space can be brought to the Yang-Mills
form is quite non-trivial. In order to arrive to such
a formulation one uses crucially the self-duality available
in four dimensions \cite{Abhay}. Thus, one might
suspect that the quantization techniques from TQFT that
use the connection field as the main variable are limited
only to gravity in three and four dimensions.

There is, however, one case that seems to contradict this
negative conclusion: the case of the usual $\rm SO(4)$
first-order formulation of gravity in four dimensions.
As we have
mentioned above, this model can be written as a $\rm SO(4)$
BF theory with additional constraints guaranteeing that the B
field comes from the frame field. This formulation
serves as the starting point for the quantum model
of Barrett, Baez and Crane, which does treat this
theory as a deformation of the BF theory.
On the other hand, the canonical formulation of this
theory is known to contain second class constraints,
solving which one does not seem to arrive
to a phase space of the Yang-Mills type
\cite{Banados}. Thus, this theory provides us with
a puzzle: on one hand, treating it covariantly as
BF theory with constraints one can quantize it
as a deformation of the topological BF theory,
on the other hand one does not expect the methods
from TQFT to work because the phase space of this
theory is not that simple as that of Yang-Mills
theory. While we do not know any simple resolution
of this puzzle, it seems from details of the quantum
theory that it uses the self-duality in some
clever way and thus goes around the problem with
the second class constraints of the canonical
formulation.

This fact, yet to be understood in full details,
opens door to a possibility of applying
the ``topological'' quantization procedure to
gravity theories in higher dimensions, hoping that the covariant quantization
will be able to go around the problem with
second class constraints that are known to be
present also in this case.
The first step that one has
to take towards this goal is to reformulate a
higher dimensional gravity theory as a BF theory
with constraints. The main aim of this paper is
to show that such a formulation is indeed possible.
In the second part of the paper we shall study in
some details the corresponding ``spin foam'' quantum theory.

\bigskip

Our results can be summarized as follows.
First, in section \ref{sec:Class},
we show that in any dimension gravity can be written as an
${\rm SO}(D)$, or ${\rm SO}(D-1,1)$, BF theory subjects to quadratic,
non-derivative constraints on the $B$ field.
Namely we prove that
gravity in $D$ dimensions can be described by the
following action functional:
\begin{equation}\label{action}
S[A,B,\Phi] = \int_{\cal M} {\rm Tr} (B\wedge F) +
{1\over 2} {\rm Tr} (B\wedge \Phi(B)).
\end{equation}
Here $A$ is an ${\rm SO}(D)$ -for spacetime of euclidean signature, or
${\rm SO}(D-1,1)$  for spacetime of Minkowskian signature- connection field.
The $B$ field is a Lie algebra
valued $(D-2)$-form and  $\Phi$ is a  Lagrange multiplier
field that can be contracted in a special way (see below)
with the $(D-2)$-form $B$ to
produce a Lie-algebra valued 2-form, denoted by $\Phi(B)$ in
(\ref{action}).
Let us emphasize that the Lagrange multiplier term of the action
is quadratic in $B$ in any dimension.
The precise form of this term will be given below.
As we show in the next section,
varying this action with respect to the Lagrange multipliers,
one obtains equations that guarantee that the $B$ field comes
from a frame field $e$:
\[
B=	*(e\wedge e).
\]
For such $B$, the action (\ref{action}) is
just the usual action for gravity in the first-order formulation.
This means that the theory described by (\ref{action}) is
indeed equivalent to gravity, in the sense that all solutions
of Einstein's theory are also solutions of (\ref{action}).

The second part of our paper is devoted to quantum theory. We
study a quantization of the theory described by (\ref{action})
along the lines of Refs. \cite{BC,Baez,SF}. This quantization
procedure, which can be called a ``spin foam'' quantization,
will be summarized in some details in section \ref{sec:Quant}.
For now, let us just note that in this quantization the $B$ field is
promoted to a derivative operator acting on the so-called spin networks.
The quadratic constraints become constraints on representations and
intertwiners labelling the spin networks.
Representations satisfying these constraints will be
called, following \cite{Baez}, {\it simple}.
In four dimensions simple representation of
$\rm SO(4) \approx SU(2) \times SU(2)$ were found \cite{BC,Baez} to be
the ones of the type $(j,j)$, that is, the ones
carrying the same spin under the left and right copies of $\rm SU(2)$.
In section \ref{sec:Quant} we
find all possible simple representations in any dimension.
Surprisingly, it turns out that the simple
representations are in certain precise sense
the simplest possible representations
of the gauge group. We find that in any dimension these representations
are labeled just by a single parameter.
Also, in any dimension, we construct an intertwiner
satisfying the intersection constraints.

\section{Classical theory}
\label{sec:Class}

This section is devoted to an analysis of the classical theory.
We will first present the action in several equivalent
formulations and then prove that it is equivalent to
the standard Einstein-Hilbert action. In subsection
\ref{sec:gauge} we discuss in details the issue of
dependence between the constraints. In this section
the gauge group can be taken to be either that corresponding
to Euclidean signature or to Lorentzian: all our proofs
are independent of this. For definiteness, we will
work with the Euclidean version, in which the
gauge group is ${\rm SO}(D)$, for this is what is
used in the quantum part of our paper.

\subsection{The action}
\label{sec:action}

The action for gravity in the BF formulation is a
functional of the $B$ field, the connection form $A$, and
Lagrange multipliers $\Phi$. There are two equivalent
formulations, which are both worth mentioning. In the
first formulation, which is more customary in the
context of BF theories, the $B$ field is thought of as
a Lie algebra valued $(D-2)$-form. In the second formulation
one uses the metric-independent Levi-Civita density to
construct from this $(D-2)$-form
a densitized rank two antisymmetric covariant tensor, which we will
call a bivector. We first present the
action in this second formulation, for it looks exactly the
same in any dimension $D\geq 4$.
Thus, we start by writing $B$ as a bivector $\B^{\mu\nu}_{ij}$,
where Greek characters are the spacetime indices, latin
letters are the internal
indices, and a single tilde over the symbol of $B$ represents the fact
that its density weight is one.

The action of the theory is then given by:
\begin{equation}\label{action-up}
S[A,\B,\ut{\Phi}] = \int d^D x~ \B^{\mu \nu}_{ij} F^{ij}_{\mu \nu}
+ {1\over 2} \ut{\Phi}_{\mu \nu \rho \sigma}^{ijkl} \B^{\mu \nu}_{ij}
\B^{\rho \sigma}_{kl}.
\end{equation}
The action is a functional of an ${\rm SO}(D)$ gauge field $A^{ij}_{\mu}$,
bivector fields $\B^{\mu \nu}_{ij}$, and Lagrange multiplier
fields $\ut{\Phi}_{\mu\nu\rho\sigma}^{ijkl}$.
This action is generally covariant: the bivector
fields scale as tensor densities of weight one, while the
multipliers scale as densities of weight minus one, which
is represented by a single tilde below the symbol `$\Phi$'.

In order to ensure the relation to gravity, 
the multiplier field $\ut{\Phi}_{\mu \nu \rho
\sigma}^{ijkl}$ must be such that it is completely anti-symmetric in
one set of indices, and its anti-symmetrization on the other set of
indices vanishes. There is a freedom, however, on which
set of indices the anti-symmetrization is taken to vanish.
It turns out to be more convenient for the quantum theory to
choose the anti-symmetrization on the spacetime indices
to vanish. This is the choice we make. Let us emphasize,
however, that from the point of view of the classical
theory the two possibilities are completely equivalent in
the sense that they are both enough to guarantee the simplicity of the $B$
field (for a generic, non-degenerate field $B$).

The postulated properties of the Lagrange multiplier
field $\Phi$ imply that it is of the form:
\[
\ut{\Phi}_{\mu \nu \rho \sigma}^{ijkl} =
\epsilon^{[m] ijkl} \ut{\Phi}_{[m] \mu\nu\rho\sigma},
\]
where $\epsilon^{[m] ijkl}$ is the totally
anti-symmetric form on the Lie algebra,
$[m]$ is a completely anti-symmetric
cumulative index of length $D-4$, and
$\ut{\Phi}_{[m] \mu\nu\rho\sigma}$ is a new Lagrange multiplier field,
which we, by abuse of notation, also call $\Phi$. This new
Lagrange multiplier field also has  density weight minus one.
The field
$\ut{\Phi}_{[m] \mu\nu\rho\sigma}$ has a property that its
anti-symmetrization on the spacetime indices vanishes:
\begin{equation}
	\label{anti}
\ut{\Phi}_{[m] [\mu\nu\rho\sigma]} = 0.
\end{equation}
Using this new set of Lagrange multipliers the action
can be written as:
\begin{equation}\label{action1}
S[A,B,\Phi] = \int d^D x~ \B^{\mu \nu}_{ij} F^{ij}_{\mu \nu}
+ {1\over 2} \ut{\Phi}_{[m] \mu\nu\rho\sigma}
\epsilon^{[m] ijkl}
\B^{\mu \nu}_{ij} \B^{\rho \sigma}_{kl}.
\end{equation}

Let us now give another way the action (\ref{action1}) can be
written, using the representation of the $B$ field as a $(D-2)$-form.
This is more standard in the context of BF theories. Using the
definition of the bivector $\B^{\mu\nu}$,
\begin{equation}
\B^{\mu\nu}_{ij} = {1\over 2! (D-2)!}
\tilde{\epsilon}^{\mu\nu\beta_1\ldots\beta_{D-2}}
B_{\beta_1\ldots\beta_{D-2}\,ij},
\end{equation}
one can easily check that the action (\ref{action1}) can be
rewritten as
\begin{eqnarray}\nonumber
S[A,B,\Phi] =  {1\over 2! (D-2)!} \int d^D x~
B_{\beta_1\ldots\beta_{D-2}\,ij} F_{\mu\nu}^{ij}
\tilde{\epsilon}^{\beta_1\ldots\beta_{D-2}\mu\nu}
+ \\
{1\over 2}
B_{\beta_1\ldots\beta_{D-2}\,ij} \Phi_{\mu\nu}^{ij}(B)
\tilde{\epsilon}^{\beta_1\ldots\beta_{D-2}\mu\nu},
\end{eqnarray}
where we have introduced a new two-form field $\Phi(B)$
with values in the Lie algebra. In the index notation
it is given by:
\begin{equation}
\Phi_{\mu\nu}^{ij}(B) :=
\ut{\Phi}_{\mu\nu\rho\sigma}^{ijkl} \B_{kl}^{\rho\sigma}.
\end{equation}
Thus, in the abstract notations,
one can write the action as
\begin{equation}\label{action-down}
\int_{\cal M} {\rm Tr} (B\wedge F) + {1\over 2}
{\rm Tr} (B\wedge \Phi(B)).
\end{equation}
Thus, there are two equivalent formulations of the theory. One
can use the formulation in terms of forms, given by (\ref{action-down}),
or the formulation in terms of bivectors, given by (\ref{action-up}).
In what follows, we will mostly use the formulation in terms
of bivectors.

Variation of the action (\ref{action1}) with respect to $\Phi$ gives the
following equations:
\begin{equation}\label{constr}
\epsilon^{[m] ijkl}
\B^{\mu \nu}_{ij} \B^{\rho \sigma}_{kl} =
\tilde{\epsilon}^{[\alpha] \mu\nu\rho\sigma}
\tilde{c}_{[\alpha]}^{[m]}
\end{equation}
for some coefficients $\tilde{c}_{[\alpha]}^{[m]}$. Here
$[m], [\alpha]$ are cumulative anti-symmetric indices of
length $D-4$, Lie algebra and spacetime ones correspondingly.
As one can see, when equations (\ref{constr}) are satisfied, the
coefficients $\tilde{c}_{[\alpha]}^{[m]}$ are given by:
\begin{equation}\label{constr'}
\tilde{c}_{[\alpha]}^{[m]} = {1\over (D-4)! 4!}
\epsilon^{[m] ijkl}
\B^{\mu \nu}_{ij} \B^{\rho \sigma}_{kl}
\ut{\epsilon}_{[\alpha] \mu\nu\rho\sigma}.
\end{equation}

The bivector field $\tilde{B}$ can be viewed
as a linear map from the space of spacetime
two-forms to the space of densitized internal two-forms:
$\B_{ij}(\theta) \equiv \B^{\mu \nu}_{ij} \theta_{\mu \nu}.$
We will say that $B$ is generic (or non-degenerate) if this map
is invertible.

It is clear that when $B$ comes from a frame
field $e$, $B$ identically satisfies (\ref{constr}).
The following theorem states that the reverse is true.
\begin{theo}
	In dimension $D > 4$ a generic $B$ field satisfies the constraints
	(\ref{constr}) if and only if it  comes from a frame field.
	In other words, a non-degenerate $B$ satisfies the constraints
	(\ref{constr}) if and only if there exist $e^{\mu}_{i}$ such that:
\begin{equation}
\B^{\mu\nu}_{ij} = \pm|e| e^{[\mu}_i e^{\nu]}_j,
\end{equation}
where $|e|$ is the absolute value of the determinant
of the matrix $e^\mu_i$.
\end{theo}
The condition $D>4$ is there because in four dimensions, under the same
assumptions, there is another solution  (see \cite{Laurent}) given by:
\begin{equation}
	\label{topo}
	\B^{\mu\nu}_{ij} = \pm |e| \epsilon_{ij}^{\quad\!\! kl}
	e^{[\mu}_k e^{\nu]}_l.
\end{equation}
Thus, our theorem, in particular, claims that 
this other solution appears only in four dimensions.

\bigskip

\noindent{\bf Proof:} The constraints (\ref{constr}) can be conveniently
subdivided
into the following categories:
\begin{eqnarray}
\hbox{simplicity:} &\qquad&
\B^{\mu\nu}_{[ij} \B^{\mu\nu}_{kl]} = 0 \qquad \mu,\nu \,
\hbox{distinct} \\
\hbox{intersection:} &\qquad&
\B^{\mu\nu}_{[ij} \B^{\nu\rho}_{kl]} = 0 \qquad \mu,\nu,\rho \,
\hbox{distinct} \\
\hbox{normalization:} &\qquad&
\B^{\mu\nu}_{[ij} \B^{\rho\sigma}_{kl]} =
\B^{\mu\rho}_{[ij} \B^{\sigma\nu}_{kl]} \qquad \mu,\nu,\rho,\sigma \,
\hbox{distinct}
\end{eqnarray}
The reason for this
terminology has to do with the conditions imposed by the various
constraints.

In appendix \ref{sim} we prove the following two propositions.
The first proposition states that
imposing the simplicity condition on a non-zero two-form $B_{ij}$ is
equivalent to demanding that the two-form is simple, or, in other
words, that it factors as the outer product of one-forms:
\begin{equation}
B_{[ij} B_{kl]} = 0 \hskip 0.15in
\Leftrightarrow \hskip 0.15in B_{ij} = u_{[i}
v_{j]}. \label{Plue}
\end{equation}
Note that we have omitted the density weight of $B_{ij}$ in the
above expression. For the
discussion that follows, where we treat $B_{ij}$ as a Lie-algebra
two-form, the density weight of $B$ is irrelevant.

The second proposition states that
the intersection condition on a pair of
simple two-forms ensures that they share
a common one-form factorizing both of them:
\begin{equation}
B_{[ij} {B'}_{kl]} = 0 \hskip 0.15in
\Leftrightarrow \hskip 0.15in B_{ij} = u_{[i}
v_{j]} \hskip 0.1in \hbox{and} \hskip 0.1in {B'}_{ij} =
v_{[i} w_{j]}.
\end{equation}
Moreover, the common factor $v_i$ is  uniquely determined up to scaling
when $B$ and $B'$ are not  proportional to each other.
In case $B$ and $B'$ are proportional to each other,
the above statement trivially holds, but the common form
$v_i$ is not determined uniquely: any linear
combination of it with the other one-form is also a
common form.

Let us now discuss the meaning of the  normalization
condition.
Imposing the normalization condition on two pairs of simple two-forms,
each pair of which is constructed by taking different outer
products of the same 4 one-forms, fixes the relative
normalization of the two two-forms. In other words,
given four simple two-forms
\begin{eqnarray}\nonumber
B_{ij} &=&  N u_{[i} v_{j]}\\ \nonumber
{B'}_{ij} &=& N' w_{[i} z_{j]}\\
{B''}_{ij} &=& N'' u_{[i} w_{j]}\\ \nonumber
{B'''}_{ij} &=& N''' z_{[i} v_{j]},
\end{eqnarray}
the conditions
$$B_{[ij} {B'}_{kl]} = {B''}_{[ij} {B'''}_{kl]}$$
imply that $N N' = N'' N'''$ as long as the four vectors are linearly
independent.

Let us now see what these assertions imply for our theory.
First, consider a
set of two-forms $B^{12}_{ij}, B^{13}_{ij}, \ldots
B^{1D}_{ij}$.  According to the simplicity relations, each of
these two-forms factors into one-forms, and according to the
intersection relations each pair shares a unique common factor.
Note that our assumption that $B$ is generic implies
that all two-forms $B^{\mu \nu}_{ij}$ are non-zero and that
$B$'s are not proportional to each other.
Let $v_i$ be the non-zero one-form shared by $B^{12}_{ij}$ and $B^{13}_{ij}$;
$w_i$ be the one-form shared by $B^{13}_{ij}$
and $B^{14}_{ij}$; $u_i$ be the one-form shared by $B^{14}_{ij}$
and $B^{12}_{ij}$.  Then there are three possibilities:
the  one-forms $u_i, v_i, w_i$ span a linear space of rank
(i) 3; (ii) 2; (iii) 1.
Let us consider each case separately.

Case (i). Since $u_i$ and $v_i$
are distinct one-forms that both divide $B^{12}_{ij}$,
this two-form is given by a product of $u_i, v_i$:
$B^{12}_{ij} = c u_{[i} v_{j]}$.  Likewise, we can
express the remaining two bivectors completely in terms of our three
vectors. Thus, we have:
\begin{eqnarray}\nonumber
B^{12}_{ij} &=& c u_{[i} v_{j]} \\
B^{13}_{ij} &=& c' v_{[i} w_{j]} \\ \nonumber
B^{14}_{ij} &=& c''w_{[i} u_{j]}.
\end{eqnarray}

Case (ii). Let us assume, without loss of generality,
that $v_i = w_i$.  Then this
vector divides all $B^{12}, B^{13}$  and $B^{14}$, but $u_i$ divides
both $B^{12}$ and $B^{14}$. So $B^{12}$ and $B^{14}$ are proportional, which
is excluded for a generic $B$.

Case (iii). Let us assume, without loss of generality,
that $u_i= v_i = w_i$. Then, from the definitions, we see that this
vector must divide all three bivectors. Thus, we can write
\begin{eqnarray}\nonumber
B^{12}_{ij} &=&  u_{[i} p_{j]} \\
B^{13}_{ij} &=&  u_{[i} q_{j]} \\ \nonumber
B^{14}_{ij} &=&  u_{[i} r_{j]}
\end{eqnarray}
for some suitable vectors $p_i$, $q_i$, $r_i$.

In four dimensions, the case (i) was associated with the
so-called topological sector (\ref{topo})
(see \cite{Laurent} for a discussion on this sector),
while the case (iii) was associated with the gravity sector.
In dimension higher than four, however, the case (i) cannot occur
since we have, for instance, the two-form $B^{15}_{ij}$ to
reckon with.  This two-form must have a factor in common with the
three two-forms considered previously. In the  case (i) , there is no
factor in common between the three two-forms. So the only possibility
is that $B^{15}$ is proportional to for instance  $B^{12}$ which is
not possible for a generic $B$.
Thus, we are forced to the case (iii), in which,
if we assume that the three bivectors are
distinct, the only common factor is $u_i$. We then conclude that $u_i$
divides $B^{15}_{ij}$ as well.  Continuing this reasoning, we
see that $u_i$ must divide all two-forms $B^{1 \nu}_{ij}$.

Repeating the above arguments with different values for spacetime
indices, we conclude that there exist one-forms ${e'}^1_i \ldots
{e'}^D_i$ such that ${e'}^\mu_i$ divides $\B^{\mu
\nu}_{ij}$ for any $\nu$.  If we are in the generic case, where these
vectors are pairwise distinct, this then implies that $\B^{\mu
\nu}_{ij} = (e') k^{\mu\nu} {e'}^{[\mu}_{i} {e'}^{\nu]}_{j}$, where
$k^{\mu\nu}$ are some coefficients symmetric in $\mu \nu$, and $(e')$ is the
determinant of the matrix ${e'}^\mu_i$, which is included to
give the right density weight to $B$. To find the coefficients
$k^{\mu\nu}$ we have to use the normalization constraints. From these
constraints,  we conclude that
$k^{\mu\nu} k^{\rho\sigma} = k^{\mu\rho} k^{\sigma\nu}$.
This relation implies that $k^{\mu\nu} = \pm c^{\mu} c^{\nu}$, for some
vectors $c^\mu$. Indeed, for $k^{\mu\nu}$ not equal to zero, there
exists one-form $n_\mu$ such that $k^{\mu\nu} n_\mu n_\nu = \pm 1$.
Multiplying the above relation by $n_\mu n_\nu$ we get 
$k^{\mu\nu} = \pm c^{\mu} c^{\nu}$, where $c^\mu = k^{\mu\nu} n_\nu$. 
Thus, if we rescale our vectors ${e'}^\mu_i$ as
\[
e^\mu_{i} =
|{c^1\cdots c^D}|^{-{1\over D+2}}
c^\mu {e'}^\mu_{i},
\]
then we have $\B^{\mu \nu}_{ij} =\pm  e e^{[\mu}_{i} e^{\nu]}_{j}$.
In odd dimensions we can always absorb the minus sign by redefining
the frame $e$. But in even dimensions we have
$\B^{\mu \nu}_{ij} =\pm  |e| e^{[\mu}_{i} e^{\nu]}_{j}$.
\\$\Box$

Substituting this solution of the constraints back into the action,
we find
\begin{equation}
S[A,\B(e)] =\pm \int d^D x~ |e|
e^{[\mu}_{i} e^{\nu]}_{j} F^{ij}_{\mu \nu}
\end{equation}
which is simply the standard Palatini action in terms of the
frame field $e$. Thus, our
classical theory is indeed a reformulation of general relativity.

Note that our theorem deals only with the case the $B$ field is
non-degenerate. It would be interesting to see what the 
constraints (\ref{constr}) imply in the case $B$ is degenerate.
This is of no relevance for the classical theory, where one does
not allow degenerate metrics. However, the case of degenerate
$B$ field may be quite relevant in the quantum theory, where,
as the example of 2+1 gravity suggests, the degenerate metrics
play an important role. Thus, it would be quite interesting to
study the degenerate sectors and to analyze their quantization.
We do not address this important problem in the present paper, hoping to
return to it in the future. For an analysis of degenerate
sectors in the case of four dimensions see \cite{Mike,Ted,Jerzy}

\subsection{Gauge transformations}
\label{sec:gauge}

This subsection deals with the issue of dependence between
the constraints (\ref{constr}). We show that the fact
that the constraints are not independent implies
the presence of an additional gauge symmetry in the theory.
We also discuss the problem of finding an independent
subset of the constraints.

The action functional (\ref{action}) is invariant under three different
sets of gauge transformations.  Two of these, spacetime
diffeomorphisms and frame rotations, are familiar so we need not
discuss them here, while the third is specific to our new
formulation and arises from the fact that the constraints (\ref{constr})
are not independent in more than four spacetime dimensions.

To understand this redundancy, let us find the number
of constraints that need to be imposed to guarantee that
the $B$ field comes from a frame, and compare this
number with the number of constraints in (\ref{constr}).
In $D$ dimensions, there are $D^2$ components of $e^\mu_{i}$ and
$(C_D^2)^2=D^2(D-1)^2/4$ components of $\B^{\mu \nu}_{ij}$,
which means that we need
the number of independent constraints equal to the
difference between the above two numbers, that is,
$D^2 (D^2 - 2 D - 3)/4$. The number of constraints
we have can be calculated by looking at the
equations (\ref{constr}) that one obtains by varying the action with
respect to the Lagrange multipliers $\Phi$. The free indices
in this equation are $[m]$ and anti-symmetric pairs $(\mu, \nu)$,
$(\rho, \sigma)$. The equations are symmetric in these
pairs. Thus, the number of equations in (\ref{constr}) is equal
to the product of $C_D^4$, which is the dimension of the
index $[m]$, with the number of independent entrees in
a symmetric $C_D^2\times C_D^2$ matrix. This gives
$C_D^4 {C_D^2(C_D^2+1)\over 2}$
equations. However, some of these equations are simply
definitions of the coefficients $\tilde{c}_{[\alpha]}^{[m]}$,
see (\ref{constr'}). Thus, to get the number of
constraints imposed by (\ref{constr}), we have
to subtract from the above number the number of
components in $\tilde{c}_{[\alpha]}^{[m]}$. This,
finally, gives
\[
C_D^4 {C_D^2(C_D^2+1)\over 2} - (C_D^4)^2
\]
constraints. In the case of four dimensions, this number
equals to 20, which is exactly the number of constraints
needed to go from the $B$ field to the frame. However,
already in five dimensions this number is much larger
than the number of independent constraints that are
needed: we have 250 constraints in (\ref{constr})
with only 75 independent constraints necessary.
The bottom line is that we have more constraints than
needed for $D>4$.
Since, as we proved, there is a solution to this set of
constraints, this simply means that they are highly redundant 
for $D>4$.

Thus, the fact that not all the constraints that
follow from our action principle are independent
does not cause any problems classically. However,
this may lead to problems in the quantum theory, for example,
with the definition of the partition function.
Indeed, in the partition function we have to integrate over
the set of Lagrange multipliers
$\ut{\Phi}^{ijkl}_{\mu\nu\rho\sigma}$.
This integration leads formally to a delta distribution of the
constraint
\begin{equation}\label{c}
C^{\mu\nu\rho\sigma}_{ijkl}\equiv K^{\mu\nu\rho\sigma}_{ijkl}
- K^{[\mu\nu\rho\sigma]}_{ijkl},
\end{equation}
where
\begin{equation}\label{k}
K^{\mu\nu\rho\sigma}_{ijkl}
\equiv \B^{\mu\nu}_{[ij} \B^{\rho\sigma}_{kl]}.
\end{equation}
If the constraints are not independent,
the integration over all Lagrange multipliers $\Phi$ leads to
products of delta functions that are ill-defined.
Thus, one may worry that the partition function of the
theory is not well-defined. The standard strategy to
deal with this problem is that of gauge fixing. As we show
below, the fact that the constraints are not independent
implies that there is an additional ``gauge'' symmetry
in the theory. Gauge fixing this symmetry amounts to
finding an independent set of constraints. Below we
exhibit such an independent set.

The additional ``gauge'' symmetry present because of
the redundancy of the constraints is given by the
following transformation of the multiplier fields:
\begin{equation}\label{gauge}
\delta \ut{\Phi}^{ijkl}_{\mu\nu\rho\sigma} =
\Lambda^{ijklmn}_{\mu\nu\rho\sigma\gamma\delta} \B^{\gamma\delta}_{mn}.
\end{equation}
As we shall illustrate below, this transformation
leaves the action invariant.
Here $\Lambda^{ijklmn}_{\mu\nu\rho\sigma\gamma\delta}$ is the 
gauge parameter, which must be symmetric under an interchange of
any two of the three antisymmetric index pairs $(\mu,\nu), (\rho,\sigma), 
(\gamma,\delta)$, anti-symmetric in the internal indices 
$jklm$, and its anti-symmetrization in spacetime indices 
$\mu\nu\rho\sigma$ must vanish. In addition, it must
be symmetric in the indices $i,n$. Note also that
the density weight of $\Lambda$ must be $-2$.

In order to prove that the transformation (\ref{gauge}) 
leaves the action invariant, we first have to show  
that the following relations hold:
\begin{equation}\label{rel}
B^{\mu \nu}_{i[j} K^{\rho \sigma \gamma \delta}_{klm]n} +
B^{\mu \nu}_{n[j} K^{\rho \sigma \gamma \delta}_{klm]i} +
B^{\rho \sigma}_{i[j} K^{\gamma \delta \mu \nu}_{klm]n} +
B^{\rho \sigma}_{n[j} K^{\gamma \delta \mu \nu}_{klm]i} +
B^{\gamma \delta}_{i[j} K^{\mu \nu \rho \sigma}_{klm]n} +
B^{\gamma \delta}_{n[j} K^{\mu \nu \rho \sigma}_{klm]i}
\equiv 0,
\end{equation}
where $K^{\mu\nu\rho\sigma}_{ijkl}$ is given by (\ref{k}).
The proof of this fact is as follows. Let us pick an arbitrary 
vector $v^i$ and set $u_j := v^i B_{ij}$. We then have
$v^i K_{ijkl} = u_{[j} B_{kl]}$. Using this relation, we
can obtain the following identity:
\begin{equation}
0 = u_{[j} u_k B_{lm]} = u_{[j} K_{klm]n} v^n = 
v^i B_{i[j} K_{klm]n} v^n,
\end{equation}
which implies that 
$B_{i[j} K_{klm]n} + B_{n[j} K_{klm]i} \equiv 0$. This is almost
the above relation (\ref{rel}). More precisely, to 
obtain (\ref{rel}) we set $B_{ij} = x
B^{\mu \nu}_{ij} + y B^{\rho \sigma}_{ij} + z B^{\gamma
\delta}_{ij}$, use the relation just proved, expand, and equate 
the coefficient of $xyz$ to zero. Using the relation
(\ref{rel}) one can easily convince oneself that the
transformation (\ref{gauge}), with the postulated
symmetry properties of the gauge parameter, 
leaves the action invariant.

It turns out that the set of gauge transformations (\ref{gauge})
is complete, i.e., these are all gauge symmetries
appearing because of the redundancy of the constraints. 
In other words, first, gauge fixing this symmetry amounts
to choosing an independent subset of constraints; second,
all other constraints follow from this independent 
subset and the relations (\ref{rel}).
Let us now present an independent subset of 
constraints (\ref{constr}) that is enough to 
guarantee that B comes from a frame.
As we have said above, these independent constraints, 
plus the relations (\ref{rel}) 
imply the rest of the constraints. The following
proposition is a statement to this effect.

\begin{prop} The following subset of the constraints 
\begin{eqnarray}\nonumber
(i) \hskip 0.3in &K^{\mu\,\mu+1\,\mu\,\mu+1}_{ij 12} = 0&
\\ \nonumber
(ii) \hskip 0.3in &K^{\mu-1\,\mu\,\mu\,\mu+1}_{123i} = 0&
\\ \label{indep}
(iii) \hskip 0.3in &B^{\mu \nu}_{[ij} K^{\mu \, \mu - 1 \, \mu \,
\mu + 1}_{1]234} - B^{\mu \, \mu - 1}_{[ij} K^{\mu \, \mu +
1 \, \mu \nu}_{1]234} + B^{\mu \, \mu + 1}_{[ij} K^{\mu \nu
\mu \, \mu - 1}_{1]234} = 0& \hskip 0.3in \mu + 1 < \nu
\\ \nonumber
(iv) \hskip 0.3in &B^{\mu \nu}_{[12} K^{\nu \, \nu - 1 \, \nu \,
\nu + 1}_{i]345} - B^{\nu \, \nu - 1}_{[12} K^{\nu \, \nu + 1 \,
\mu \nu}_{i]345} + B^{\mu \, \mu + 1}_{[12} K^{\mu \nu
\nu \, \nu - 1}_{i]345} = 0& \hskip 0.3in \mu + 1 < \nu 
\\ \nonumber
(v) \hskip 0.3in &K^{12\mu \nu}_{1234} = K^{1 \mu 2
\nu}_{1234}& \hskip 0.3in \mu + 1 < \nu
\end{eqnarray}
is independent. Moreover, the above constraints, plus the
relations (\ref{rel}) imply all the constraints (\ref{constr}).
\end{prop}

\bigskip

A proof of this proposition consists of two parts. First,
as it is easy to see, the number of constraints in
(\ref{indep}) is just the right number of constraints
needed to go from the $B$ field to a frame. Indeed,
there are $D(C_{D-2}^2)$ constraints in (i), $D(D-3)$ in (ii),
$(C_D^2 - D)C_{D-1}^2$ in (iii), $(C_D^2 - D)(D-2)$ in (iv),
and, finally, $C_D^2 - D$ constraints in (v). Adding all these
numbers together one obtains exactly $(C_D^2)^2 - D^2$, which
is the number of DOF in the $B$ field minus the number
of DOF in the frame. Thus, the counting shows that
the number of constraints (\ref{indep}) is not 
larger than the number of constraints that is needed. The 
second part of the proof is to show that the constraints
(\ref{indep}) are enough to guarantee that the $B$ field
comes from a frame. This is done by expressing all
the other constraints as constraints (\ref{indep}) 
modulo the relations (\ref{rel}).
We will not give a proof of this fact here, for it 
involves a rather lengthy manipulation with the relations
(\ref{rel}). 
\\ $\Box$

\section{Quantum models}
\label{sec:Quant}

This section is devoted to quantum theory. More precisely, we
apply the so-called ``spin foam'' quantization procedure to
our theory. We first review the main steps of this procedure,
and then find results analogous to ones available in the
case of four dimensions.

\subsection{Spin foam quantization}
\label{sec:sfoam}

In the paper \cite{SF}  it was advocated that  the knowledge of the
generating functional $Z[J]$ 
\begin{equation}
Z[J] = \int {\cal D}A {\cal D}B \, 
e^{i\int_{\cal M} {\rm Tr}\left[B\wedge F + B\wedge J\right]}
\end{equation} 
of the BF theory,
hence the ability to compute  all correlation functions of the  $B$
field in the BF theory, opens a way towards understanding of
Yang-Mills theories in any dimension and of gravity in three and four
dimensions. The key point is to consider these
theories as  deformations of the BF theory.
The knowledge of $Z[J]$  leads to an understanding of these theories in
the same way as the knowledge of the generating functional of the
free scalar field leads to understanding of an interacting quantum field
theory. Similar ideas were also put forward
by Martellini and collaborators for the case
of Yang-Mills theory \cite{Martellini}.
As far as gravity is concerned, a strength of this proposal
is in the fact that the BF theory
incorporates gauge invariance and needs no background metric for its
definition, which are two features desired for a non-perturbative
treatment of gravity.

Moreover, in \cite{SF} a ``spin foam'' computation of the generating
functional was performed. The resulting  ``spin foam'' version of the
generating functional immediately gives
the ``spin foam'' quantization of any theory
considered as a constrained BF theory.
In this quantization the field $B$ is promoted to a derivative
operator (in $J$) acting on the generating functional $Z(J)$.

In the previous section we showed that  gravity in higher
dimensions can be written as BF theory with constraints.
This means that gravity in any dimension
falls within the scope of applicability of the method \cite{SF}.
Thus, there exists a spin foam model of higher-dimensional
gravity, some aspects of which we study below.
The spin foam model we obtain is a generalization of the 4D spin foam model
proposed in \cite{R,BC,Baez} to higher dimensions.

In this paper we describe only the main steps of the spin foam
quantization procedure. For details the reader may consult
Refs. \cite{BC,Baez,SF}. One starts with a decomposition of $\cal M$ into
piecewice-linear cells. For simplicity this decomposition is usually
taken to be a triangulation which we shall denote by
$\Delta$; it will be fixed
in what follows. Having a fixed triangulation, one
can compute the spin foam ``approximation'' to the
generating functional $Z[J]$ of the BF theory.
This is an approximation because it takes into
account only special distributional configurations
of the B field. However, as it was shown in \cite{SF}, this
approximation is exact for TQFT.

The result of calculation of $Z[J]$ can be described
as follows. Up to fine details related to the
way the simplex amplitudes are glued together,
$Z[J]$ can be thought of as given by a sum over
product of amplitudes -- one for each $D$-simplex. These
amplitudes depend on the  current
through a collection of group elements. The current,
being a two-form, can be integrated over the special portions
of the dual faces of $\Delta$ that are called
in \cite{R,SF} {\it wedges}.
Each wedge is in one-to-one
correspondence with a pair ($D$-simplex, $(D-2)$-simplex lying in it).
Integrating $J$ over all wedges of $\Delta$, one gets
a collection of Lie algebra elements. Exponentiating
the later one obtains a collection of group elements.
The generating functional $Z[J]$ depends on the current
through these group elements.

The simplex amplitude is obtained as follows. First,
one has to construct a special graph.
The boundary of each $D$-simplex is a $(D-1)$-dimensional
manifold triangulated by $(D-1)$-simplices. One can
construct a graph dual to this triangulation. In $D$ spacetime
dimensions, this graph will have $D+1$ vertex and $D(D+1)/2$
edges. Each vertex will have exactly $D$ edges coming to it,
or, in other words, its valency will be $D$. For each
edge, let us take the usual space $L^2(G)$ of square integrable
functions on the group. Edges of these graph are in
one-to-one correspondence with wedges introduced above,
and, thus, with the group elements constructed above from 
the current $J$. Thus, we can think of elements of $L^2(G)$
for each edge of our graph as functions of the group
elements constructed from the current. To construct the
vertex amplitude, which is a function of all $D(D+1)/2$
group elements coming from $J$, one has to choose
the so-called {\it intertwiner} for each vertex. Intertwiners
give a way to construct a function of $D(D+1)/2$ group
elements that is invariant under the action of the group.
This function is the simplex amplitude. In practice,
the simplex amplitudes are given by the so-called
{\it spin networks}, which are constructed by taking
a basis in $L^2(G)$ consisting of the matrix elements
of the irreducible representations.

Given the generating functional $Z(J)$, the computation of
the expectation values of products of $B$ field is given by derivative
operators acting on $Z(J)$.
Thus, amplitudes for gravity theory can be obtained from the
above simplex amplitudes by imposing on them certain differential
equations with respect to the current. This procedure can be
justified as a projection on the kernel of constraints
arising when one takes the path integral over the Lagrange
multipliers $\Phi$:
\begin{equation}
\int {\cal D}\Phi 
e^{i\int_{\cal M} {1\over 2} {\rm Tr}\left[B\wedge \Phi(B)\right]}
= \delta(C),
\end{equation}
where $C$ are the constraints (\ref{c}).
After finding solutions to the differential equations corresponding
to these constraints, one evaluates them on $J=0$ to
obtain amplitudes for gravity.

There are several types of constraints that one has
to impose. First, there are the so-called closure
constraints. These arise because one is considering
a set of Lie algebra two-forms $B_{ij}$, one for each
$(D-2)$-simplex, that are obtained by integrating the $B$
field, which is a $(D-2)$-form, over these $(D-2)$-simplices,
and there are linear dependences between $B_{ij}$
obtained this way. It is straightforward to solve
the differential equations corresponding to these
constraints for they simply require the simplex
amplitude to be gauge invariant.

Second, there are simplicity constraints for each $(D-2)$-simplex,
or for each two-form $B_{ij}$, which require this two-form
to be simple. This constraints can also be solved
in quantum theory. They imply that only a part of
the space $L^2(G)$ is relevant. This relevant part
can be written as a direct sum over special representations
that can be called {\it simple representations}. We find
and study some properties of these representations in the
following subsection.

Third, there are analogs of intersection constraints. In
the quantum theory these constraints appear as constraints
on intertwiners. We find a solution to these constraints
in subsection \ref{sec:spinnet}.

Finally, there is a problem of imposing analogs of
normalization constraints.
However, these are non-trivial already in the case of
four dimensions. Already in that case, there exists
a lot of confusion in the literature as to this problem.
We will not discuss it in this paper.

In this paper we will not discuss the spin foam model itself;
instead, we would like to understand what are the implications,
from  the point of view
of representation theory, of the simplicity and intersection
constraints.

\subsection{Simple representations}
\label{sec:simple}

In this section we use notations and general results on
representation
theory of ${\rm SO}(D)$ that are described in Appendices \ref{SO} and
\ref{sphe}. We refer the reader to the books \cite{Knapp,GT} for a deeper
exposition of the results stated in these two appendices.

Let us denote the basis of the Lie algebra of ${\rm SO}(D)$ by
$X_{ij}$, $i,j \in \{1,\ldots, D\}$. The commutation relations
are given by (\ref{com}).
As we have discussed in the previous section, two-forms
$B_{ij}$ are promoted in the ``spin foam'' quantization
to derivative operators acting on the generating functional.
Since Lie algebra is generated by derivatives (vector fields)
on the group, this means that $B_{ij}$ is promoted in the
quantum theory to an
element $X_{ij}$ of the Lie algebra of
${\rm SO}(D)$.

The quantum analog of the Pluecker relation (\ref{Plue}) is given by:
\begin{equation}
	X_{[ij}X_{kl]}=0, \, \forall i,j,k,l\in \{1,\ldots, D\}, \label{QP}
\end{equation}
where $[ijkl]$ means that we consider the total anti-symmetrization on
these indices.

Given a linear representation $V$ of ${\rm SO}(D)$ we say will that $V$ is a
{\it simple} representation if the quantum Pluecker relation
(\ref{QP}) is identically satisfied on $V$.

It is clear that if $V$ is simple, it decomposes into a
sum of irreducible simple representations of ${\rm SO}(D)$; so it is enough to
concentrate on irreducible simple representations.
The purpose of this subsection is to give a
complete classification of the space of simple representations
of ${\rm SO}(D)$
for all values of $D$. We find that the simple representations in any
dimension are labelled by only one positive integer. Thus, there
is a remarkable similarity between simple representations in any
dimension $D$ starting from $D=3$.

There is a natural representation of ${\rm SO}(D)$
in the space ${\cal L}^2(S^{D-1})$
of square integrable functions on the $(D-1)$-sphere. The group
action is given by:
\begin{equation}
	g \cdot \phi(x) = \phi(g^{-1} x),
\end{equation}
where $x=(x_1, \cdots,x_D)$ is a unit vector from $R^D$.
This representation is reducible: any ${\cal L}^2$
function on the sphere can be decomposed into spherical harmonics
\begin{equation}
	{\cal L}^2(S^{D-1}) = \oplus_{N=0}^{\infty} H_{N}^{(D)},
\end{equation}
where $H_{N}^{(D)}$ represents the space of harmonic homogeneous
polynomial of degree $N$ (see Appendix \ref{sphe}).
The action of the Lie algebra elements $X_{ij}$ in
the space ${\cal L}^2(S^{D-1})$ is given by:
\begin{equation}
	X_{ij}\cdot \phi(x) = x_i {\partial \phi \over \partial x_j}(x) -
	x_j {\partial \phi\over \partial x_i}(x).
\end{equation}
It is now obvious to see that the space ${\cal L}^2(S^{D-1})$, and,
therefore,
$\H_{N}^{(D)}$ gives a simple representation.
\begin{equation}
	\label{simpl}
		X_{[ij}X_{kl]}\phi = x_{[i}\partial_jx_{k}\partial_{l]} \phi
		= x_{[i}\delta_{jk}\partial_{l]}\phi +
		x_{[i}x_{k}\partial_j\partial_{l]}\phi
		=0
\end{equation}
The first equality is the  definition of the representation, the second
is obtained by commuting $x$ and $\partial$, and the third by taking into
account the anti-symmetrization on the indices.

A remarkable fact is that the spherical harmonics representations
$\H_{N}^{(D)}$ are the only simple representations of ${\rm SO}(D)$.
The following theorem is a statement to this effect:

\begin{theo}
	$V$ is an irreducible simple representation of $SO(D)$, $D\geq 4$, if
and only
	if $V$ is equivalent to one of the  representations $\H_{N}^{(D)}$.
\end{theo}

We would like to present two proofs of this fact, one
by recurrence and the other more direct.
Both these proofs use the fact that an irreducible representation  is uniquely
characterized by its highest weight. Moreover,
the highest weight characterizing $\H_{N}^{(D)}$ is $N e_1$, in the
notation of appendices \ref{SO} and \ref{sphe}.

Before we give the proofs, let us make several comments. First,
for $D=3$  there is no Pluecker relation; one can say
that all representation of ${\rm SO}(3)$ are simple. In that case
the above theorem is still valid, because the
representations $\H_{N}^{(D)}$ exhaust all representations
of ${\rm SO}(3)$. They correspond, of course, to the integer spin
representations
of ${\rm Spin}(3)\equiv {\rm SU}(2)$.

The second comment is that in dimension $D=4$, the theorem
has already been proved in \cite{BC}.
However, these authors did not realize the crucial
fact that the simple representations are
related to spherical harmonics. This is this fact that
allows us to find the generalization of simple representations
to higher dimensions. Let us see how the simple
representations of \cite{BC} are related to the ones
described in the above theorem. In $D=4$,
$X_{[ij}X_{kl]}$ is an invariant tensor (there is only one such
tensor because in dimension four there is a unique  totally anti-symmetric
tensor of rank four).
The value of this tensor on the representation
$\Lambda(n_1,n_2) = e_1 (n_1 +n_2) /2 + e_2 (n_1 -n_2)/2$
(see  appendix \ref{SO}) is given by   $n_1(n_1+2) - n_2(n_2+2)$.
In that case the quantum Pluecker relation reads $n_1=n_2=N $, so
simple representations of ${\rm SO}(4)$
are given by the highest weight $ \Lambda = N e_1$, $N$ being
a positive integer.
This is precisely the highest weight of the representation
$\H_{N}^{(4)}$.
Let us now give the proofs for a general dimension $D$.

\bigskip

\noindent{\bf Proof 1} As we have just discussed, the theorem
holds for dimensions
$D=3, 4$. Thus, to prove the theorem in any dimension,
it is enough to show that from the fact that it holds in
dimension $D$ it follows that it holds in $D+1$. Thus, we assume
that representations $\H_{N}^{(D)}$ are the only simple
irreducible representations of $\SO(D)$ and show that
from this assumption it follows that $\H_N^{(D+1)}$ are
the only simple irreducible representations of $\SO(D+1)$.
We prove this by constructing an
embedding of $\SO(D)$ into $\SO(D+1)$ and then showing that the
pullback of simple representations of $\SO(D+1)$ under
this embedding is a simple representation of $\SO(D)$.
As the last step of the proof we show
that the only irreducible representations of $\SO(D+1)$
that have the property that their pullback contains the
representations $\H_N^{(D)}$ are the representations
$\H_N^{(D+1)}$.

Let us construct an embedding of $\SO(D)$ into $\SO(D+1)$.
Since ${\rm SO}(D+1)$ is the group of rotation of $D+1$
dimensional vectors, ${\rm SO}(D)$ can be obtained as the subgroup
fixing the vector $(0,\ldots,0,1)$.
This gives an embedding of ${\rm SO}(D)$ into
${\rm SO}(D+1)$, which we denote by
\[
\phi: \SO(D) \rightarrow \SO(D+1).
\]
If $X_{ij}$, $i,j \in \{1,\ldots, D\}$ is a basis of $so(D)$, then
the action of the embedding $\phi $ on the Lie algebra is given by
$\phi(X_{ij}) = X_{ij}$ for $i,j \in \{1,\ldots, D\}$.
If $V$ is a representation of ${\rm SO}(D+1)$ we can define its
pullback $\phi^*(V)$ using the embedding $\phi$.
In other words, when $V^{(D+1)}_\Lambda$ is a representation
of $\SO(D+1)$ of highest weight $\Lambda$, $g \in {\rm SO}(D)$ and
$ v\in V^{(D+1)}_\Lambda $, then $g\cdot \phi^*(v) = \phi(g) \cdot v$.
Thus, $\phi^*(V^{(D+1)}_\Lambda) $ is a representation of
$\SO(D)$. However, this representation is not necessarily
irreducible, but it certainly contains, when decomposed
into a sum of irreducible representations,
the irreducible  subrepresentation of highest weight $\phi^*(\Lambda)$.
This is because the embedding $\phi$ maps positive roots onto
positive roots and $\phi^*(v_\Lambda)$ is the highest weight for
$\SO(D)$.

Let us now take as $ V^{(D+1)}_\Lambda $ a simple representation of
$\SO(D+1)$. Then the pullback $\phi^*(V^{(D+1)}_\Lambda) $ is a simple
representation of $\SO(D)$. Indeed, considering a basis of $so(D)$,
we have:
\begin{equation}
	X_{[ij}X_{kl]}\cdot \phi^*(V^{(D+1)}_\Lambda) \equiv
	\phi(X_{[ij}) \phi(X_{kl]})\cdot V^{(D+1)}_\Lambda =
	X_{[ij}X_{kl]}\cdot V^{(D+1)}_\Lambda =0.
\end{equation}
Here the first equality is the definition of the pullback of a
representation, in the second we used the definition of the
embedding, and the
third uses the hypothesis that $ V^{(D+1)}_\Lambda$ is simple.

This result means that $\phi^*(\Lambda)$ is the highest weight
of a simple representation of $\SO(D)$. We have assumed
that all highest weight simple representations of $\SO(D)$
come from weights $N e_1$. In other words,
$\phi^*(\Lambda)$ must equal $N e_1$ for some $N$.
We would like to show now that the only highest
weights $\Lambda$ of $\SO(D+1)$ satisfying this
property are the ones corresponding
to the representations $H_N^{(D+1)}$, i.e., $\Lambda= N e_1$.
To see this we have to consider two cases: (i) $D=2n$;
(ii) $D=2n+1$.

(i) In this case, $\SO(2n)$ and $\SO(2n+1)$ have the same rank $n$,
and $\phi^*$ is the identity operator, $\phi^*(e_i) =e_i$.
Thus, $\phi^*(\Lambda)= N e_1 $
implies that $\Lambda= N e_1 $.

(ii) In this case, $\SO(2n+1)$ has rank $n$, while $\SO(2n+2)$
has rank $n+1$. Thus, $\phi$ is  the projection operator
$\phi^*(e_i) =e_i$, $ i\leq n$ and  $\phi^*(e_{n+1}) = 0$.
Therefore, $\phi^*(\Lambda)= N e_1$
implies that $\Lambda= N e_1  + k e_{n+1}$ for some $k$.
But $\Lambda$ is a highest weight of $\SO(2n+1)$. This means (see
appendices B and C) that $ \Lambda = N_1 e_1 + N_2 e_2 + \cdots + N_{n+1}
e_{n+1}$ where $ N_1 \geq N_2 \geq \cdots \geq N_{n+1} \geq 0$.
Thus, if $n\geq 2$, the only highest weight $\Lambda$ of  $\SO(2n+1)$
satisfying  $\phi^*(\Lambda)= N e_1 $ is given by
$\Lambda= N e_1 $.

\bigskip

\noindent{\bf Proof 2} Here is a more direct proof of
the theorem that uses the correspondence (\ref{cartan})
between the Cartan basis and the usual basis of $so(D)$.
Let us denote for $1\leq i<j \leq n $, $n= [D/2]$,
\begin{equation}
	C(i,j)= -3 X_{[2i-1,2i} X_{ 2j-1, 2j]}.
\end{equation}
If $ V^{(D)}_\Lambda $ is a simple  representation of
$\SO(D)$, then the action of $C(i,j)$ vanishes on  $ V^{(D)}_\Lambda $.
Using  (\ref{cartan}) and after some algebra
we get:
\begin{equation}
C(i,j) = H_j(H_i+1) + E_{-e_i- e_j} E_{e_i+ e_j} -
E_{-e_i+ e_j} E_{e_i- e_j}.
\end{equation}
Evaluating this expression on the highest weight vector $v_\Lambda$
we get:
\begin{equation}
 C(i,j)v_\Lambda = (\Lambda|e_j)((\Lambda|e_i) +1)v_\Lambda.
\end{equation}
The simplicity constraint implies  that
$(\Lambda|e_j)((\Lambda|e_i) +1)=0$ for all $1\leq i< j \leq n$,
which means $ (\Lambda|e_j) =0$ for all $1 <j\leq n$. Thus, the
only highest weights representations that satisfy the simplicity
constraint correspond to $\Lambda= Ne_1$.
\\$\Box$

\subsection{Simple spin networks}
\label{sec:spinnet}
In this section we deal with the quantum version of the
intersection constraints. As we discussed above, these
become equations on intertwiners. In this subsection we
show how to solve these constraints in any dimension.
We arrive at a notion of {\it simple} spin networks,
whose edges are labelled by simple representations satisfying 
the simplicity constraints, and
whose intertwiners satisfy the intersection
relations. The simple spin networks are
higher dimensional generalizations of the relativistic
spin networks of \cite{BC}. However, let us first recall
some general facts about spin networks. A more detailed account
is given in \cite{Baez}.

An $\SO(D)$ spin network is a triple $(\Gamma,\Lambda,\imath)$, where
(i) $\Gamma$ is an oriented graph, (ii) $\Lambda$ is a labelling of
each edge $e$ by an irreducible representation $\Lambda_e$
of $\SO(N)$, (iii) $\imath$
is a labelling of each vertex by an intertwiner $\imath_v$ mapping the
tensor product of incoming representations at $v$ to the product
of outgoing representations at $v$.
Let us denote by $I(\Gamma,\Lambda,v)$ the space of such intertwiners
at vertex $v$. We can then
associate to each colored graph $(\Gamma,\Lambda)$ a vector space
\begin{equation}
	\H(\Gamma,\Lambda) = \otimes_v I(\Gamma,\Lambda,v).
\end{equation}
When $\Gamma$ has only one edge and one vertex (circle),
this space is one
dimensional and is generated by the function on $\SO(N)$ given by the
character in the representation $\Lambda$.
For a general graph this space can be thought of as the space of
functionals in the variables $\Lambda_e(g_e)$ which are
invariant under gauge transformation acting at each vertex of $\Gamma$.
This is the space of spin network functionals based on the
colored graph $(\Gamma,\Lambda)$.

Given an oriented edge $e$, let us denote by
$e_-$ the vertex where $e$ starts and by
$e_+$ the vertex where $e$ ends. For any pair $(e,\pm)$,
we can define a group action on $\H(\Gamma,\Lambda)$. This is
given by the right or left multiplication:
\begin{eqnarray}\nonumber
	h^{(e,e_+)} \phi(g_{e_1}, \cdots, g_{e},\cdots, g_{e_n}) =
	\phi(g_{e_1}, \cdots,	 g_{e}h,\cdots, g_{e_n}), \\ \nonumber
		h^{(e,e_-)} \phi(g_{e_1}, \cdots, g_{e},\cdots, g_{e_n}) =
	\phi(g_{e_1}, \cdots,	h^{-1} g_{e},\cdots, g_{e_n}).
\end{eqnarray}
We denote  $X^{(e,e\pm)} $ the corresponding action of the Lie algebra
by the derivative operator.
We can now introduce the notion of {\it simple spin networks}.

We say that an $\SO(N)$ spin network $\phi \in \H(\Gamma,\Lambda)$
is {\it simple} if for all vertices $v$ and all pairs of edges
$e,e'$ meeting at $v$, the following relation is satisfied:
\begin{equation}
 	X^{(e,v)}_{[ij}X^{(e',v)}_{kl]}	\phi =0. \label{SS}
\end{equation}
This relation, for a pair $(e,e)$,
amounts to the quantum Pluecker relation (\ref{QP}) for the representation
$\Lambda_e$. This means that the edges of a simple spin network are
labelled by simple representations, that is, are
colored by one integer $N_e$ which characterize the simple
representations $\H^{(D)}_{N_e}$.
The remaining conditions (\ref{SS}) for distinct pairs of edges meeting
at a vertex $v$ are conditions on the intertwining operator used at
the vertex $v$.

Let $e_1, \ldots, e_n$ be the incoming edges and
$e'_1,\ldots, e'_p$ be the outgoing edges at the vertex $v$.
Then $\H^{(D)}_{N_{e_i}},\H^{(D)}_{N_{e'_j}}$ are simple
representations associated with edges meeting at $v$.
An intertwiner from the tensor product of incoming
simple representations to the product of outgoing ones
is given by a multi-linear map
 \begin{equation}
 	I(P_1,\cdots,P_n, {\bar Q}_1,\cdots, {\bar Q}_p),
 \end{equation}
where $ P_i$ ($ Q_j$ respectively) are harmonic homogeneous polynomials of
degree $ N_{e_i}$ ($N_{e'_j}$ respectively), and ${\bar Q}$ denotes the
complex conjugate of $Q$.
The intertwining property reads
\begin{equation}
	I( P_1,\cdots,P_n, {\bar Q}_1,\cdots, {\bar Q}_p) =
	I(g\cdot P_1,\cdots,g\cdot P_n, g\cdot {\bar Q}_1,\cdots, g\cdot
{\bar Q}_p),
\end{equation}
and an example of the relation (\ref{SS}) is given by:
\begin{equation}
I(X_{[ij}\cdot P_1, X_{kl]}\cdot P_2,\cdots,P_n,
{\bar Q}_1,\cdots, {\bar Q}_p) = 0
\end{equation}

There exists a very simple and beautiful solution of these constraints.
This solution was discovered for the case $D=4$ by \cite{BC}.
However, in that work, it was written in a
rather cumbersome way as a sum over a product of intertwiners of $SU(2)$.
Moreover, the proof that the intertwiner satisfies
the intersection constraints used heavily the fact that the universal
covering of
$SO(4)$ can be written as the product $SU(2)\times SU(2)$.
This uses the duality available in $D=4$, which makes this dimension
very special.
Thus, it was not at all clear that this solution could be generalized
to higher dimensions, where there is no notion of duality.
The solution we give shows that the central notion,  allowing the
construction to work, is not  self-duality, but the fact that simple
representations are realized in the space of polynomials on the sphere
$S^{D-1}$.

Let $ P_i$ ($Q_j$ respectively) be harmonic homogeneous polynomial of
degree $N_{e_i}$ ($N_{e'_j}$ respectively) and consider the
following intertwiner between
\[
	\otimes_{i=1}^{n}\H^{(D)}_{N_{e_i}}\, {\mathrm and}\,
	\otimes_{j=1}^{p} (\H^{(D)}_{N_{e'_j}})
\]
given by
\begin{equation}\label{Int}
 I_{n,p}(P_1,\cdots,P_n, {\bar Q}_1,\cdots, {\bar Q}_p) =
 \int_{S^{D-1}} d\Omega(x) \,P_1(x) \cdots P_n(x)
 {\bar Q}_1(x) \cdots {\bar Q}_p(x).
\end{equation}
Here $d\Omega$ denotes the invariant measure on the unit sphere
$S^{D-1}$. For definiteness we choose the normalization of this
measure such that
\[
\int_{S^{D-1}} d\Omega(x) =1
\]

The fact that $I_{n,p}$ is an invariant intertwiner can be
easily seen using the invariance of the measure, integration by parts
and the Leibniz rule. This is, in a sense, the simplest possible
intertwiner one can imagine. Note that the above intertwiner
in the case $D=3$  is the usual intertwiner of $\SO(3)$
that is constructed from Clebsch-Gordan coefficients.
It is remarkable that such a simple entity gives a
simple intertwiner:
\begin{theo}
 $I_{n,p}$ satisfies the relation (\ref{SS}).
\end{theo}

\bigskip

\noindent{\bf Proof:}
Let us consider a vertex with $n$ incoming and $p$ outgoing edges.
Let us choose any two of the incoming edges, which we denote by
$e_1$ and $e_2$. Let us denote by $F$ the product of all
polynomials $P, Q$ except $P_{e_1}, P_{e_2}$:
$F = \prod_{i\neq 1,2}P_{e_i} \prod_{j} Q_{e_j}$.
We then have to prove that the following quantity
\begin{equation}
	\label{int}
	X^{e_1}_{[ij}X^{e_2}_{kl]}I_{n,p}= \int_{S^{D-1}}
	d\Omega X_{[ij} \cdot P_{e_1}  X_{kl]} \cdot P_{e_2} F
\end{equation}
is zero. We can use the following identity:
\begin{equation}
	2 X_{[ij} \cdot P_{e_1}  X_{kl]} \cdot P_{e_2}=
	 X_{[ij} \cdot  X_{kl]} \cdot( P_{e_1}  P_{e_2}) -
	   (X_{[ij}X_{kl]} \cdot  P_{e_1})  P_{e_2}-
	     P_{e_1} (X_{[ij} X_{kl]}\cdot P_{e_2}),
\end{equation}
The last two terms are zero because $P$'s are homogeneous
polynomials. Similarly, the first term is zero because the
product of homogeneous polynomials is also a homogeneous
polynomial. The other intersection relations can be
proved analogously.
\\$\Box$

One question that remains is the question of uniqueness of
the intertwiner we constructed. However, this question is
non-trivial already in the case of four dimensions.
In the case of $D=4$ there exists an argument
\cite{Reis} that shows that this intertwiner is
the only one satisfying all the constraints. Thus, it
may be the case that the intertwiner we found is unique
in any dimension. It would be interesting to find
a proof of this conjecture. We leave this issue to 
further research.

\section{Discussion}
\label{sec:Disc}

We have seen that, in many aspects, the BF formulation of higher-dimensional
gravity is analogous to the four-dimensional case. Indeed, as
in four dimensions, one must add to the usual BF action constraints
that are quadratic in the $B$ field and that guarantee that it
comes from the frame field. We also saw that the quantum
spin foam models in higher dimensions are quite similar to
their four-dimensional cousin. Strikingly, in any dimension representations
that appear are labelled by just one parameter, the structure
of the intertwiner that is used to built the model is
quite similar to that in four dimensions. Let us emphasize
that this similarity between the case of four dimensions
and higher dimensional theories is by itself an interesting
and unexpected result. Indeed, as we discussed in the Introduction,
it is tempting to believe that the four-dimensional case
is special, for there the self-duality is available. Our results
indicate that the case of four dimensions is not that special.

There are, however, several differences between the case
of four dimensions and higher-dimensional gravity that
are worth mentioning. First, unlike the
four-dimensional case, in higher dimensions it is much harder
to single out the independent constraints. In four dimensions
the number of Lagrange multipliers that appear in the action
is equal to the number
of independent constraints. In higher dimensions we were not
able to find a covariant formulation with this property: the
number of Lagrange multipliers appearing in the action
(\ref{action}) is much larger than the number of independent
constraints. We were able, however, to find a description of independent
constraints, see the subsection of \ref{sec:Class} on gauge
transformations, but not in any covariant way. Thus, unlike
the four-dimensional case, we don't have an action principle
with the number of Lagrange multipliers appearing equal to
the number of independent constraints. This does not seem,
however, to cause any problems, either classically or
quantum mechanically. Classically one finds complicated
relations (\ref{rel}) between the constraints appearing from
varying the action (\ref{action}), and this is manifested
by the appearance of gauge symmetries discussed in
section \ref{sec:Class}. However, all relations
together, although not independent,
do imply that the $B$ field comes from a frame field.
One might worry that the dependence of constraints may
cause problems quantum mechanically.
However, as we saw, it doesn't seem to be the case, at least in the spin foam 
context. As we have seen in the last section, it was
possible to impose the simplicity and intersection
constraints on spin foam by explicitly constructing the intertwiners
satisfying these constraints.

The second important difference between $D=4$
and the higher-dimensional cases is the absence of the
topological sector. As we saw in subsection \ref{sec:action},
in higher dimensions there is only one type of solutions
of the simplicity constraints, in contrast to two
different types in the case of four dimensions: the case
(i), according to the classification of the subsection
\ref{sec:action}, can exist only in four dimensions,
where it leads to the topological sector. This is an
interesting feature of higher-dimensional gravity, for
it means that one does not have to worry about a possible
interference between the two sectors when they are
both present in the quantum theory. Also, unlike the four-dimensional
case, there is no worry that the spin foam quantization gives
a quantization of the topological sector, not gravity: simply because
there is no topological sector anymore. Of course, one still has to
worry about the issue of ``two signs'', arising in the
solution of the constraint equations. But this comes
about even in the simplest case of three dimensions,
where the two types of solutions can interfere in
the quantum theory and make the problem of finding
the ``gravitational'' sector of the theory very difficult,
see \cite{3DVolume} for a discussion of this problem.

Let us now discuss implications of our results for the
problem of quantization of gravity. We have discussed
some aspects of the ``spin foam'' quantum model of gravity 
for all $D > 4$. As we saw, these models turn out to 
be quite similar to the four-dimensional model. There
are, however, many problems with this model even in the
case of $D=4$, of which the main one is probably
that one does not know how to glue the simplex amplitudes
together to form the amplitude of the whole triangulated
manifold. The other problem is that we do not know yet how to implement the
normalization conditions in the quantum theory.
Thus, the quantum theory presented in this paper is far from 
giving a correct quantization of gravity.

Our results, however, have another interesting implication
for the problem of quantum gravity. Our results imply
that gravity, as well as Yang-Mills
theory in any dimension, can be thought of as an
``interacting'' BF theory. Indeed, the action of
this theories can be rewritten as that of BF theory
plus an additional term quadratic in the $B$ field,
which can be thought of as the ``interaction'' term.
This means that the problem of quantization of
both Yang-Mills and gravity theories in any
dimension to a large extent reduces to the problem of finding
the generating functional $Z[J]$ of the BF theory:
\[
Z[J] = \int {\cal D}A {\cal D}B \,
e^{i\int_{\cal M} {\rm Tr}(B\wedge F) + {\rm Tr}(B\wedge J)},
\]
where $J$ is the current two-form. Indeed, because
actions for the both theories can be represented in
the form BF action plus quadratic term in $B$, all
correlation functions of these theories (in $B$ field) can be
found by appropriately differentiating the
generating functional $Z[J]$ with respect to $J$.
Thus, $Z[J]$ is a universal object, a knowledge of which
in a particular spacetime dimension
to a large extent means the knowledge of both
Yang-Mills and gravity theories in that dimension.
This way of approaching the problem of four-dimensional quantum gravity
was advocated in \cite{SF}. The results of our paper
mean that this strategy can also be applied to higher
dimensional theories. Let us also mention that a partial
progress along the lines of finding $Z[J]$ was achieved in \cite{SF},
where we found a ``spin foam'' approximation to
this generating functional in any dimension.

Let us conclude by pointing out another interesting
implication of our results. In four spacetime dimensions,
the use spin foam models
was to a large extent motivated by results of
the canonical approach to quantum gravity \cite{LQG}:
the known four-dimensional spin foam models are
intimately related to the loop canonical quantization
of gravity. We have found that the spin foam model
formulation of quantum gravity is not limited to
four dimensions. Thus, our results point towards an
interesting possibility that there exists an analog
of canonical connection quantization of gravity
in any dimension. Work is currently in progress
on trying to find such a formulation.

\bigskip

{\bf Acknowledgements}: We are grateful to L. Smolin for a discussion. 
This work was supported in part by the NSF
grant PHY95-14240 and by the Eberly research funds of Penn State. 
K.K. was supported in part by a Braddock fellowship from Penn State.

\appendix

\section{Pluecker relations}
\label{sim}

Relations
which enforce that a multivector factors as an anti-symmetrized product
of vectors arise in the geometry of subspaces of linear spaces, and
are known as Pluecker relations. In this appendix, we shall review
and demonstrate some relevant facts of algebraic geometry concerning
those relations.

As the first step, we shall consider the case of a single
two-form $B_{ij}$, and show that the necessary and sufficient condition for
it to be an anti-symmetrized product of one-forms is the following:
\begin{equation}
	B_{[ij} B_{kl]} = 0.
\end{equation}
Before showing that this condition implies factorization, we will
note that it is equivalent to the following weaker condition:
$${1\over 2} B_{i[j} B_{kl]} = B_{ij} B_{kl} + B_{ik} B_{lj} + B_{il}
B_{jk} = 0.$$
To see that this is the case, we simply write out the six terms
appearing in the complete anti-symmetrization on the indices $ijkl$, and
note that each of the three possible ways of choosing a pair of two
indices appears twice. Let us now show that the above condition
implies that the two-form $B_{ij}$ factors as a product of
one-forms. If $B_{ij}$ is not identically zero, then we
can find vectors $a^i$ and $b^j$ such that $B_{ij} a^i b^j = 1$.
Let us now define one-forms $u_i$ and $v_i$ as $u_i = B_{ij} a^j$
and $v_i = B_{ij} a^j$. Then, using the above identity, we obtain
$$u_i v_j - v_i u_j = (B_{ik} B_{jl} - B_{il} B_{jk}) a^k b^l =
B_{ij} B_{kl} a^k b^l = B_{ij}.$$
Thus, this proves the simplicity of $B_{ij}$ by explicitly
constructing the two one-forms that divide it.

Next, given two simple two-forms, let us find a condition that they
have a non zero common factor.  First, assuming that this is the case, we have:
$$B_{ij} = u_{[i} v_{j]} \hskip 0.4in \hbox{and} \hskip 0.4in
{B'}_{ij} = v_{[i} w_{j]}.$$
We then see that $\lambda B_{ij} + \mu {B'}_{ij}$
must be a non zero simple bivector for any value of the constants
$\lambda,\mu$.
Using the simplicity criterion that we have already proved above,
this will be the case  if the relation
$$B_{[ij} {B'}_{kl]} = 0$$
is satisfied. This is the relation we looked for. Let us now
show that this relation is also a sufficient condition for
two two-forms to have a common factor.
To show this we shall produce this common factor explicitly.
Consider the following entity:
$$B_{i[j} {B'}_{kl]}.$$
If the two bivectors appearing in the above expression were
proportional, then, because they are simple, this
expression would vanish identically.
If they are not proportional, it is possible to choose vectors $a^i,
b^i$, and $c^i$ such that the expression
$$v_i := B_{i[j} {B'}_{kl]} a^j b^k c^l$$
differs from zero.  The fact that this one-form is a factor of
$B_{ij}$ follows immediately from:
$$B_{[ij}v_{k]} = B_{[ij} B_{k][l} {B'}_{mn]}a^l b^m c^n = 0,$$
where we have used the fact that $B_{ij}$ satisfies the
simplicity constraint.
To see that our vector also divides the other bivector, we note that
there is another expression for $v_i$. The Pluecker
relation reads:
$$0 = B_{[ij} {B'}_{kl]} =
B_{i[j} {B'}_{kl]} + {B'}_{i[j} B_{kl]}.$$
This means that the roles of the two bivectors in our formula are
interchangeable. Hence $v_i$ is a factor of both bivectors.
This vector is unique up to rescaling, for the two two-forms are
distinct.

\section{Some facts about SO(N) and its representation theory}
\label{SO}

We will denote by $X_{ij}$, $i,j \in \{1,\ldots, D\}$ generators
of Lie algebra ${\mathrm SO}(D)$. They satisfy the following
commutation relations:
\begin{equation}
	\label{com}
	[X_{ij},X_{kl}] = \delta^{ik}X_{jl} - \delta^{il}X_{jk}
	-\delta^{jk}X_{il} + \delta^{jl}X_{ik}
\end{equation}

Let us consider the Cartan representation of this Lie algebra.
There are two cases to consider: (i) $D=2n+1$; (ii) $D=2n$.
In the first case the
corresponding Dynkin diagram is $B_n$; in the second case it is $D_n$.

The Cartan subalgebra ${\cal H}$ of $\SO(2n+1)$ is generated by
$H_k=iX_{2k-1 2k}, k=1,\ldots, n$. We denote by $e_k$ the generators of
the dual of ${\cal H}$, $e_k(H_j) = \delta_{kj}$.

Let us denote by $\Delta \in {\cal H}^*$ the root space of $so(D)$.
We have
\begin{eqnarray}\nonumber
\Delta &=& \{ \pm e_i \pm e_j , {\mathrm with }\, 1\leq i<j \leq n\}
\cup \{ \pm e_i , 1\leq i \leq n\}, \qquad {\mathrm for }\, D = 2n+1.
\\ \nonumber
\Delta &=& \{ \pm e_i \pm e_j , {\mathrm with}\, 1\leq i<j \leq n\},
\qquad {\mathrm for }\, D = 2n.
\end{eqnarray}
The Cartan basis $H_i,E_{\pm \alpha}$, $ i\in \{1,\ldots,n\}$,
$\alpha \in \Delta$ is related to the basis $X_{ij}$ by
\begin{eqnarray} \label{cartan}
	H_i&=&  i X_{2i-1 2i} \\
	E_{e_i +e_j}&=&  {1\over 2i} [ X_{2i-1,2j-1} -i X_{2i-1,2j}
	                          -i X_{2i, 2j-1} -X_{2i,2j}], \\
    E_{-e_i-e_j} &=& {1\over 2i} [ X_{2i-1,2j-1} +i X_{2i-1,2j}
	                          +i X_{2i, 2j-1} -X_{2i,2j}], \\
	E_{e_i- e_j} &=& {1\over 2i} [ X_{2i-1,2j-1} +i X_{2i-1,2j}
	                          -i X_{2i, 2j-1} -X_{2i,2j}], \\
	E_{-e_i+e_j} &=&{1\over 2i} [ X_{2i-1,2j-1} -i X_{2i-1,2j}
	                          +i X_{2i, 2j-1} -X_{2i,2j}]. \\
\end{eqnarray}
where $1\leq i< j\leq n$.

The simple roots are given by:
\begin{eqnarray}
\alpha_1= e_1-e_2,\,\ldots,\, \alpha_{n-1} = e_{n-1} -e_{n},\, \alpha_n =
e_n,\,\quad  {\mathrm for }\, D= 2n+1, \\
\alpha_1= e_1 - e_2,\, \ldots,\, \alpha_{n-1} = e_{n-1} - e_{n}, \,
\alpha_n = e_{n-1} + e_n,\,\quad {\mathrm for }\, D= 2n.
\end{eqnarray}

One of the central theorems in the theory of Lie groups states that
the irreducible representations of the covering group of $\SO(D)$,
i.e. ${\rm Spin}(D)$, are in one-to-one correspondence with the dominant
integral weights, that is, weights of the form:
\[
\Lambda = \sum_{i=1}^{n}n_i \lambda_i,
\]
where $n_i$ are positive integers, $\lambda_i$ are the Dynkin weights
satisfying ${2 (\lambda_i, \alpha_j) / (\alpha_j, \alpha_j) }= \delta_{ij}$
and  $n=[D/2]$ ($[\cdot]$ is the integral part).
$\Lambda$ denotes the highest weight of the corresponding
irreducible representation.

When expressed in the basis given by the weights $e_i, 1\leq i \leq n$, the
highest weights labelling representations of ${\rm Spin}(D)$ are given by:
\begin{eqnarray}
	\Lambda(n_1, \cdots, n_n) = ( n_1+ \cdots + n_{n-2} + n_{n-1}
+{n_{n}\over 2}) e_1&  \\
    + \cdots +  (n_{n-2} + n_{n-1} + {n_{n}\over 2}) e_{n-2} +
	  (n_{n-1} + {n_{n}\over 2}) e_{n-1} +
	  ( {n_{n} \over 2}) e_n, \,&  \rm{ for}\, D = 2n+1  \\
	\Lambda(n_1, \cdots, n_n) = ( n_1+ \cdots + n_{n-2} + {n_{n-1}
+n_{n}\over 2}) e_1 & &\\
    + \cdots +  (n_{n-2} + {n_{n-1} +n_{n}\over 2}) e_{n-2} +
	  ({n_{n-1} +n_{n}\over 2}) e_{n-1} +
	  ( {n_{n} -n_{n-1}\over 2}) e_n, \,\,&
	  \rm{ for}\, D=2n
 \end{eqnarray}

The irreducible representations of $\SO(D)$
are in one-to-one correspondence with
the irreducible representations of ${\rm Spin}(D)$
that satisfy the restriction:
(i) $n_n$ is an even integer for $D=2n+1$;
(ii) $n_{n-1} + n_{n}$ is an even integer for $D=2n$.

\section{Harmonic polynomial representations of SO(N)}
\label{sphe}

Let $V_N^{(D)}$ be the space of complex-valued homogeneous
polynomials of degree $N$ on $R^{D}$. Then
$\SO(D)$ acts on this space by $ g\cdot P( x) = P(g^{-1} x)$.
This action induces the following action of the Lie algebra $\SO (D)$:
\begin{equation}
	X_{ij}\cdot P(x) = x^i {\partial P\over \partial x^j}(x) -
	x^j {\partial P\over \partial x^i}(x).
\end{equation}
The basis of weight vectors of $V_N^{(D)}$ is given by:
\begin{eqnarray}
	(x_1+ix_2)^{k_1}(x_1-ix_2)^{l_1}\cdots (x_{2n-1}-ix_{2n})^{l_1},&
	{\mathrm for } \, D=2n \\
		(x_1+ix_2)^{k_1}(x_1-ix_2)^{l_1}\cdots
(x_{2n-1}-ix_{2n})^{l_1}
		x_{2n+1}^{k_0},&{\mathrm for } \, D=2n+1,
\end{eqnarray}
where $ N = \sum_j k_j + \sum_il_i$.
This basis diagonalizes the action of the Cartan subalgebra generated
by $H_i= i X_{2i-1 2i} $. Hence, the weights of $V_N^{(D)}$ are given
by:
\begin{equation}
	\sum_{i=1}^{j=n} (k_i -l_i) e_i.
\end{equation}
The highest weight of $V_N^{(D)}$ is $N e_1$ and its dimension is
$ C^{N+D-1}_N $.

This representation is, however, not irreducible. To see this,
let us consider the Casimir
\[
C = {1\over 2} \sum_{ij} X_{ij}X_{ij}.
\]
Its action on $V_N^{(D)}$ is given by:
\begin{equation}
	 C\cdot P(x) = N(N+D-2) P(x) + |x|^2 \Delta P(x),
\end{equation}
where $\Delta = (\partial_1)^2 + \cdots +(\partial_D)^2$ is the Laplacian.
Thus, an invariant subspace of this action is the
subspace $\H^{(D)}_N$ of harmonic homogeneous polynomials on $R^D$.
This space is known to be an irreducible representation of $\SO(D)$. Thus,
$\H^{(D)}_N$ can be equivalently characterized as the irreducible
representation of highest weight $\Lambda=N e_1$.
The dimension of $\H^{(D)}_N $ can be deduced from the dimension of
$V^{(D)}_N$ using the following relations
\begin{eqnarray}
{\mathrm dim}{\H^{(D)}_N} ={\mathrm dim}{V^{(D)}_N} -{\mathrm
dim}{V^{(D)}_{N-2}}, \\
{\mathrm dim}{\H^{(D)}_N} =  \sum_{k=1}^{N} {\mathrm dim}{\H^{(D-1)}_k}.
\end{eqnarray}
From the first relation we deduce that
\begin{equation}
	{\mathrm dim}{\H^{(D)}_N} = {(N+D-3)!(2N+D-2) \over N! (D-2)!}.
\end{equation}
The second equality tells us how $\H^{(D)}_N$, viewed as a
representation of $SO(D-1)$, decomposes as a sum of irreducible
representations.

\end{document}